\documentclass[12pt,preprint]{aastex}
\usepackage{graphicx,natbib}
\usepackage{amsmath,amsthm}
\def\p{\partial}
\def\eg{{\it e.g.\,}}
\defcitealias{Barnes09}{BQLR}
\defcitealias{richardson00}{Richardson et al. 2000}
\defcitealias{Porco08}{Porco et al. 2008}

\shorttitle{Hill's Equations}
\begin{document}
\title{A Symplectic Integrator for Hill's Equations}
\author{Thomas Quinn}
\affil{Department of Astronomy, University of Washington}
\affil{Box 351580, Seattle, WA, 98195}
\email{trq@astro.washington.edu}
\author{Randall P. Perrine}
\affil{Department of Astronomy, University of Maryland}
\affil{Computer \& Space Sciences Building, Stadium Drive College
  Park, MD, 20742}
\author{Derek C. Richardson}
\affil{Department of Astronomy, University of Maryland}
\affil{Computer \& Space Sciences Building, Stadium Drive College
  Park, MD, 20742}
\and
\author{Rory Barnes}
\affil{Department of Astronomy, University of Washington}
\affil{Box 351580, Seattle, WA, 98195}

\begin{abstract}
Hill's equations are an approximation that is useful in a number of
areas of astrophysics including planetary rings and planetesimal
disks.  We derive a symplectic method for integrating Hill's
equations based on a generalized leapfrog.  This method is implemented
in the parallel $N$-body code, {\em PKDGRAV} and tested on some
simple orbits.  The method demonstrates a lack of secular changes in
orbital elements, making it a very useful technique for integrating
Hill's equations over many dynamical times.  Furthermore, the method
allows for efficient collision searching using linear extrapolation of
particle positions.
\end{abstract}


\section{Introduction}



There are a number of situations in planetary dynamics that require
the exploration of near-circular orbits.  Current topics of interest
in this category include planetary rings \citep{WisdomTremaine88} and
planet formation \citep[hereafter BQLR]{Tanga04,Barnes09}.  In this circumstance
the equations of motion can be linearized about the circular orbit
as was first done by \cite{Hill1878} to
study the lunar orbit.  That is, the motions of bodies are described
with respect to a Cartesian frame that is in uniform circular motion about a
central body, and excursions from the center of the frame are small
compared to the distance to the central body.  In the
absence of perturbations, the resulting 
Hill's equations describe simple
epicyclic motion, and can also be
used for disk dynamics \citep{GoldreichLyndenBell65,JulianToomre66}
and the escape of stars from globular
clusters \citep{Heggie2001}.

A noticeable disadvantage of Hill's equations for numerical integration
is that they contain a velocity-dependent force.  Simulation codes for
large $N$-body simulations, \citep[e.g.][]{GADGET,Wadsley04} typically use
the leapfrog integration scheme, which is second order, symplectic,
and easy to implement.  The leapfrog scheme can be modified to take
velocity-dependent forces into account and still retain second order,
as is done for Smoothed Particle Hydrodynamics (SPH); however, this destroys
its symplectic nature.

The power of symplectic integrators is rooted in the property that any
truncation error can be represented as a perturbing Hamiltonian.
Hence for sufficiently small step size, the numerical system has
conserved quantities similar to the
integrals of motion of the physical system.  That is, the numerical
integration is an exact solution to an approximate Hamiltonian.  This
property is
particularly important when following systems for many dynamical times
such as the long-term evolution of the Solar System, or investigating
the stability of extrasolar planetary systems.  In these situations,
if the integrator introduces secular changes in the actions, the
dynamics being investigated can be fundamentally changed.  Hence symplectic
integrators are widely used in such
investigations
\citep{HolmanWisdom93,LevisonDuncan94,Malhotra95,LeePeale02,RiveraLissauer00}.
Both planetary rings and planetesimal dynamics are systems that evolve
over large numbers of dynamical times, and therefore may also benefit from
the use of symplectic integrators.
 
A symplectic integrator for Hill's equations was introduced by
\cite{Heggie2001} in the context of escape of stars from globular
clusters.  However, in that work, the integrator is not actually put to use;
instead, the orbits were calculated using a Hermite integrator.
The symplectic integrator was expressed as an implicit set of
equations which, however, could be solved explicitly.
As shown below, a canonical transformation can significantly simplify the
Hamiltonian, and therefore simplify the resulting integrator.

\cite{SahaTremaine92} introduced a formalism 
\citep[also see][]{WisdomHolman91} for
deriving symplectic integrators of a generalized leapfrog type by
separating the
Hamiltonian into parts that can be integrated exactly and then using
commutator algebra to combine these solutions into a symplectic
solution to the full problem.  \cite{Quinn97} showed how this
technique could be used for cosmological
simulations that involve a time-dependent Hamiltonian. 
This formalism has also been used to construct higher-order integrators
\citep{LaskarRobutel01,ChambersMurison00}, integrators that handle
close encounters \citep{Chambers99,DuncanLevisonLee98}, and integrators
that efficiently integrate problems with a large dynamic range
\citep{SahaTremaine94,McNeilNelson09}.
Here we will apply the technique to Hill's
equations.  In section 2 present the Hamiltonian formulation of Hill's
equations from which in section 3 we derive a symplectic integrator
suitable for
use in a large $N$-body code.  In section 4, we describe its
implementation in the {\em PKDGRAV} $N$-body code \citep{Stadel01},
explicitly stating the algorithm for timestepping a simulation, and in
section 5, we
perform tests appropriate for the application of planetesimal
dynamics in the early Solar System.  Section 6 contains a short
discussion and summary.

\section{Hamiltonian Formulation}

The Lagrangian for Hill's equations in the orbital plane is
\begin{equation}
{\cal L}  = \frac{1}{2}\left[(\dot x - \Omega y)^2 + (\dot y + \Omega
  x)^2\right] + \frac{1}{2} \Omega^2(2 x^2 - y^2) - \Phi(x,y),
\end{equation}
\citep{Heggie2001} where $x$ and $y$ are, respectively, the distances
perpendicular to and along the
direction of rotation from the center of a frame in circular motion with angular speed $\Omega$.  $\Phi$ is the potential due to other forces, e.g.,
interactions with other particles.  In all that follows, we will
neglect the motion in the $z$ direction since it is trivial to
integrate in the standard way.

Lagrange's equations
give the standard Hill's equations of motion,
\begin{eqnarray}
\label{eq:motionx}
\ddot{x} - 2\Omega\dot{y} - 3\Omega^2 x & = & -\frac{\partial\Phi}{\partial x}\\
\label{eq:motiony}
\ddot{y} + 2\Omega\dot{x} & = & -\frac{\partial\Phi}{\partial y}.
\end{eqnarray}
In this form, the presence of the velocity-dependent terms requires a
modification to the leapfrog method such that a predicted velocity is
used in the estimate of the final acceleration.  This maintains second
order, but is obviously not time reversible and destroys the
symplectic nature of leapfrog.  However, as we show below, a
symplectic integrator can be derived for this system.

To construct a symplectic integrator, we first derive the Hamiltonian form of
the equations of motion.
From their definitions, the canonical momenta are
\begin{eqnarray}
p_x & \equiv & { \p {\cal L} \over \p \dot x} = \dot x - \Omega y\\
p_y & \equiv & {\p {\cal L} \over \p \dot y} = \dot y + \Omega x
\end{eqnarray}
and the Hamiltonian is
\begin{equation}
{H}(x, y, p_x, p_y) = \frac{p_x^2}{2} + \frac{p_y^2}{2} + \Omega (y p_x - x p_y) -
\frac{1}{2} \Omega^2(2 x^2 - y^2) + \Phi(x,y). \label{eq:ham1}
\end{equation}
Now consider a new set of canonical coordinates, $(X, Y, P_x, P_y)$,
derived from the generating function
\begin{equation}
S_2(x, y, P_x, P_y) = xP_x + yP_y - \Omega x y.
\end{equation}
The rules of canonical transformations then give
\begin{equation}
p_x = \frac{\p S_2}{\p x} = P_x - \Omega y; \quad
p_y = \frac{\p S_2}{\p y} = P_y - \Omega x,
\end{equation}
and
\begin{equation}
X = \frac{\p S_2}{\p P_x} = x; \quad Y = \frac{\p S_2}{\p P_y} = y.
\end{equation}
In terms of the original positions and velocities, these new canonical
coordinates are $x$, $y$, $P_x = \dot x$, and $P_y = \dot y + 2 \Omega
x$.  The Hamiltonian in these coordinates is
\begin{equation}
{H}(x, y, P_x, P_y) = {P_x^2 \over 2} + {P_y^2 \over 2} - 2
\Omega x P_y + {\Omega^2 x^2 \over 2} + \Phi(x,y),
\end{equation}
a somewhat simpler form than equation (\ref{eq:ham1}).

Hamilton's equations of motion are therefore
\begin{eqnarray}
\dot x & = & P_x\\
\dot y & = & P_y - 2 \Omega x\\
\dot P_x & = & 2 \Omega P_y - \Omega^2 x - {\p \Phi \over \p x}\\
\dot P_y & = & - {\p \Phi \over \p y}.
\end{eqnarray}
From these equations, it is obvious that $P_y$ is constant in the
absence of perturbing forces.  This 
is equivalent to the conservation of angular momentum, and leads to
the conserved quantity, $\sum_i {P_y}_i$, in a many-particle system.
In particular, $\sum_i {P_y}_i$ is conserved in a collision between
particles \citep{WisdomTremaine88}, which will be useful when calculating
collision outcomes (see below).  However, for periodic boundary
conditions, $P_y$ will change as a particle crosses the boundary in $x$
because of the shear across the box.  Nevertheless the time-averaged
total $P_y$ should be constant for a system that does not have a net
motion in the $x$ direction \citep{WisdomTremaine88}.
Also from the equations of motion in this form it is clear why Hill's
equations are
easy to integrate numerically.  If the guiding center of the motion is
at $x = 0$, then $P_y = 0$ for all time in the absence of perturbations,
and the motion reduces to a harmonic oscillator with frequency $\Omega$.

\section{Symplectic Integrators}

Normally one can create a symplectic integrator by separating the
Hamiltonian into exactly integral parts as in \cite{SahaTremaine92},
but the presence of a velocity-dependent force makes this approach
nontrivial in the case of Hill's problem.  The Hamiltonian can be split
as follows,
\begin{equation}
H = \underbrace{\frac{1}{2}(P_x^2 + P_y^2)}_{H_{FP}}  +
\underbrace{ -2 \Omega x P_y + {\Omega^2 x^2 \over 2} + \Phi(x,y)}_{H_{MDF}},
\end{equation}
where $H_{FP}$ is the free particle Hamiltonian and $H_{MDF}$ is
considered to be the (momentum-dependent) ``force'' term.  The
$H_{FP}$ Hamiltonian is easily solved and is just the motion of a
particle with constant velocity:
\begin{eqnarray}
\label{eq:drift}
x(t_0 + \tau) & = & x(t_0) + \tau P_x(t_0)\notag\\
y(t_0 + \tau) & = & y(t_0) + \tau P_y(t_0)\\
P_x(t_0 + \tau) & = & P_x(t_0)\notag\\
P_y(t_0 + \tau) & = & P_y(t_0),\notag
\end{eqnarray}
where $\tau$ is the timestep and $t_0$ is the initial time.
However, the equation of motion
corresponding to the $H_{MDF}$ Hamiltonian can not be solved easily.
This is
because for this part of the Hamiltonian $y$ is not constant ($\p H_{MDF}/\p P_y \ne 0$), so one must
evaluate the force
along a trajectory determined by $\dot y = - 2 \Omega x$ to solve for
$P_y$, and then use this to solve for
$P_x$.  This would prove intractable in a large simulation.

Instead, let us separate the Hamiltonian into a mixed term,
$H_{\it Mix}$,
and a momentum-independent force term, $H_{\it MIF}$, as follows:
\begin{equation}
H = \underbrace{\frac{1}{2}(P_x^2 + P_y^2 ) -2 \Omega x P_y}_{H_{\it Mix}}  +
\underbrace{{\Omega^2 x^2 \over 2} + \Phi(x,y)}_{H_{\it MIF}}.
\end{equation}

$H_{\it MIF}$ is easily integrated to give the equations of motion,
\begin{eqnarray}
\label{eq:kickx}
P_x(t_0 + \tau) & = & P_x(t_0) - \tau \left(\Omega^2 x(t_0) + \left. \p
\Phi \over \p x \right|_{t_0} \right) \\
P_y(t_0 + \tau) & = & P_y(t_0) - \tau \left. \p \Phi \over \p y
\right|_{t_0}.\notag
\end{eqnarray}
The mixed Hamiltonian, $H_{\it Mix}$, gives the equations of motion,
\begin{eqnarray}
\dot x & = & P_x\\
\dot y & = & P_y - 2\Omega x\\
\dot P_x & = & 2\Omega P_y\\
\dot P_y & = & 0.
\end{eqnarray}
These can be integrated exactly as follows.  $P_y$ is a constant so
$P_y(t) = P_y(t_0)$.  $P_x(t)$ can now be solved.  $x(t)$ can be solved
once $P_x(t)$ is known, and finally $y(t)$ can be solved since we
know $P_y(t)$ and $x(t)$.  We therefore have
\begin{eqnarray}
\label{eq:drift2}
x(t_0 + \tau) & = & x(t_0) + \tau P_x(t_0) + \tau^2 \Omega P_y(t_0) \notag\\
y(t_0 + \tau) & = & y(t_0) + \tau (P_y(t_0) - 2 \Omega x(t_0)) - \tau^2
\Omega P_x(t_0) - \tau^3 \frac{2}{3} \Omega^2 P_y(t_0) \\
P_x(t_0 + \tau) & = & P_x(t_0) + \tau 2 \Omega P_y(t_0) \notag\\
P_y(t_0 + \tau) & = & P_y(t_0).\notag
\end{eqnarray}

These can be used to construct a second-order symplectic integrator
exactly analogous to leapfrog by applying
equations (\ref{eq:kickx}) for half a timestep, equations
(\ref{eq:drift2}) for a
full timestep, and equations (\ref{eq:kickx}) for another half timestep. 
If we were simply
integrating
force equations this is straightforward to implement in a large $N$-body
code.  However, in the case of planetesimal and planetary ring
dynamics, collisions between particles need to be detected.  Current
collision detection algorithms rely on the
position updates being linear in time \citep{richardson00}, and certainly
not cubic in time
as in the above.


In an attempt to simplify the mixed equations of motion, let us
separate $H_{\it Mix}$ even further into the free particle Hamiltonian, $H_{FP}$, and the
cross term, $H_C = - 2 \Omega x P_y$.  That is,
\begin{equation}
H = \underbrace{\frac{1}{2}(P_x^2 + P_y^2)}_{H_{FP}}  +
\underbrace{ -2 \Omega x P_y}_{H_C} + \underbrace{{\Omega^2 x^2 \over
    2} + \Phi(x,y)}_{H_{\it MIF}}.
\end{equation}
The cross Hamiltonian is easily integrable giving
\begin{eqnarray}
\label{eq:crossx}
x(t_0 + \tau) & = & x(t_0)\\
\label{eq:crossy}
y(t_0 + \tau) & = & y(t_0) - \tau 2 \Omega x(t_0)\\
\label{eq:crosspx}
P_x(t_0 + \tau) & = & P_x(t_0) + \tau 2 \Omega P_y(t_0)\\
\label{eq:crosspy}
P_y(t_0 + \tau) & = & P_y(t_0).
\end{eqnarray}
Therefore, a first-order symplectic scheme presents itself as
follows.  1) Update the momenta using equations (\ref{eq:kickx}).
2) Update $P_x$ using equation (\ref{eq:crosspx})
and the $P_y$ from step 1. 3) Update the $y$ positions of the
particles according to equation (\ref{eq:crossy}).
If collisions are being considered, they are
searched for in this step.  4)
Perform a standard position update using the free particle Hamiltonian
(equation \ref{eq:drift}), again
searching for collisions.  In our implementation, steps 3 and 4 are
combined into a single position update that includes the collision search.

The construction of a second-order scheme follows using the formalism
of \cite{SahaTremaine92}.
If we refer to the evolution of phase space for a time $\tau$ under
the Hamiltonians
$H_{\it MIF}$, $H_{FP}$, and $H_C$ as $MIF(\tau)$
(eq. \ref{eq:kickx}), $FP(\tau)$ (eq. \ref{eq:drift}),
and $C(\tau)$ (eq. \ref{eq:crossx}-\ref{eq:crosspy})
respectively, then the combination of operators
$MIF(\tau/2) C(\tau/2) FP(\tau) C(\tau/2) MIF(\tau/2)$ will evolve the
system for a timestep $\tau$ with second-order accuracy.   That is,
the error Hamiltonian will be of order $\tau^3$ or
higher \citep{SahaTremaine92}.

\section{Implementation in an $N$-body code}


To test the usefulness of this formulation we have implemented the
above integration algorithm in the parallel gravity code {\em
  PKDGRAV} \citep{Stadel01} as part of the technique for solving an
$N$-body system in a patch corotating in a Kepler
potential \citepalias{richardson00,Porco08,Barnes09}.  This code uses a standard
form of leapfrog, where the velocities are first updated by a half step,
a {\em Kick}, then the positions are updated by a full step, a {\em
  Drift}, and finally the velocites are given a second half step {\em
  Kick}.  Only minor changes were
needed to implement the above second-order algorithm. 
The most straightforward way to modify the algorithm is to change the 
{\em Kick} routine
so that it only includes the terms present in the $H_{\it MIF}$
Hamiltonian, and modify the {\em Drift} routine to include the
operations of equations (\ref{eq:crossx}) through (\ref{eq:crosspy})
as well as the standard {\em Drift} of the positions.  However, the existing
{\em Drift} routine in {\em PKDGRAV} is complicated by the handling of periodic
boundary conditions and the search for collisions, so we instead
rearranged the operations so that the {\em Drift} remains a simple linear
extrapolation of the positions with constant velocities.

In detail, the modifications are as follows.  After the opening {\em Kick}
routine updates the velocities according to $H_{\it MIF}$, it calculates
the canonical momentum, $P_y$, and updates $\dot x$ (which is equivalent
to $P_x$) using equation (\ref{eq:crosspx}).  However, $\dot y$ is
updated to be the sum of the contributions of the free particle
Hamiltonian, $H_{FP}$, the cross term (\ref{eq:crossy}) applied at the
beginning of the
{\em Drift} (for time $\tau/2$), and the cross term applied at the end of
the {\em Drift} (also for a time $\tau/2$):
\begin{equation}
\label{PyToYdot}
\dot y = P_y - \left(\frac{1}{2}\right) 2 \Omega x -
\left(\frac{1}{2}\right) 2 \Omega (x + \tau \dot x).
\end{equation}
(Also see equation \ref{kickopenl} below.)
This $\dot y$ along with $\dot x$ can now be used to linearly
update the positions according to both ``Cross'' operators and the
free particle
operator using an essentially unmodified {\em Drift} routine.  Finally, the
closing {\em Kick} again uses equation (\ref{eq:crosspx}) to update $\dot
x$, sets $\dot y$ to be $P_y - 2 \Omega x$, and updates the
velocities according to $H_{\it MIF}$.
The above algorithm requires
storage for a new attribute, $P_y$, for each particle.

One modification to the {\em Drift} routine involves the handling of
periodic boundary conditions.
Often, Hill's equations are integrated with periodic
boundaries in the $x$ and $y$ directions.  If a particle exits
the computational volume in the, \eg, positive $y$ direction, then it is
replaced by a particle with the same velocity and $x$ coordinate
on the negative $y$ boundary.  Handling the $x$ boundaries
is a little more complicated because of the shear across the patch: a
particle in a circular orbit on the outer boundary is moving slower in
the $y$ direction than a particle on a circular orbit on the inner
boundary by an amount $\frac{3}{2} \Omega \Delta x$, where $\Delta x$
is the width of the patch.  This implies an increase in the
$P_y$ of the particle of $\frac{1}{2} \Omega \Delta x$.  This
corresponds to the fact that in a Kepler potential, the circular
velocity of an orbit decreases outwards, while the angular momentum of
a circular orbit increases outwards.  Hence when the {\em Drift} routine
detects a particle has crossed the $x$ boundary, $P_y$ is changed accordingly.

The {\em Drift} routine also performs collision detection and resolution.
Therefore if a particle's momentum is changed due to a collision, or
if a new particle is created either due to merging or fragmentation,
the canonical momentum, $P_y$, needs to be updated to reflect the change.
Solving equation (\ref{PyToYdot}) for $P_y$ provides a simple means
to calculate a new $P_y$ at any time during the drift.
For a merger, the conservation of the total $P_y$ of the bodies involved in the
collision could be used to assign a $P_y$ to the merged particle.
No other changes to {\em PKDGRAV} were needed to implement this algorithm.

In summary, a single timestep of a single particle starting from the
state $(x_n, y_n, \dot x_n, \dot y_n)$ and producing the state
$(x_{n+1}, y_{n+1}, \dot x_{n+1}, \dot y_{n+1})$ is performed as follows.
\begin{eqnarray}
\dot x_{n + 1/4} & = & \dot x_n - \frac{\tau}{2}\left(\Omega^2 x_n + \frac{\p
  \Phi(x_n, y_n)}{\p x}\right)\label{kickopenf}\\
P_y & = & \dot y_n + 2 \Omega x_n - \frac{\tau}{2}\frac{\p \Phi(x_n,
  y_n)}{\p y}\\
\dot x_{n + 1/2} & = & \dot x_{n+1/4} + \tau \Omega P_y\\
\dot y_{n+1/2} & = & P_y - \Omega x_n - \Omega(x_n + \tau \dot x_{n+1/2})\label{kickopenl}\\
x_{n+1} & = & x_n + \tau \dot x_{n+1/2}\label{driftf}\\
y_{n+1} & = & y_n + \tau \dot y_{n+1/2}\label{driftl}\\
\dot x_{n+3/4} & = & \dot x_{n+1/2} + \tau \Omega P_y\label{kickclosef}\\
\dot x_{n+1} & = & \dot x_{n+3/4} - \frac{\tau}{2}\left(\Omega^2 x_{n+1} + \frac{\p
  \Phi(x_{n+1}, y_{n+1})}{\p x}\right)\\
\dot y_{n+1} & = & P_y - 2 \Omega x_{n+1} - \frac{\tau}{2}\frac{\p \Phi(x_{n+1},
 y_{n+1})}{\p y}\label{kickclosel}
\end{eqnarray}
Although fractional subscripts are used for intermediate values of the state
variables for clarity, the only extra storage needed for this update
is for the canonical momentum, $P_y$.
Equations (\ref{kickopenf} - \ref{kickopenl}) are implemented
in the first {\em Kick} routine, and equations (\ref{kickclosef} -
\ref{kickclosel}) are implemented in the last {\em Kick} routine,
leaving a simple form for the {\em Drift}, equations (\ref{driftf}) and
(\ref{driftl}), during which boundary crossing and collision detection
is performed.

\begin{figure}
\plotone{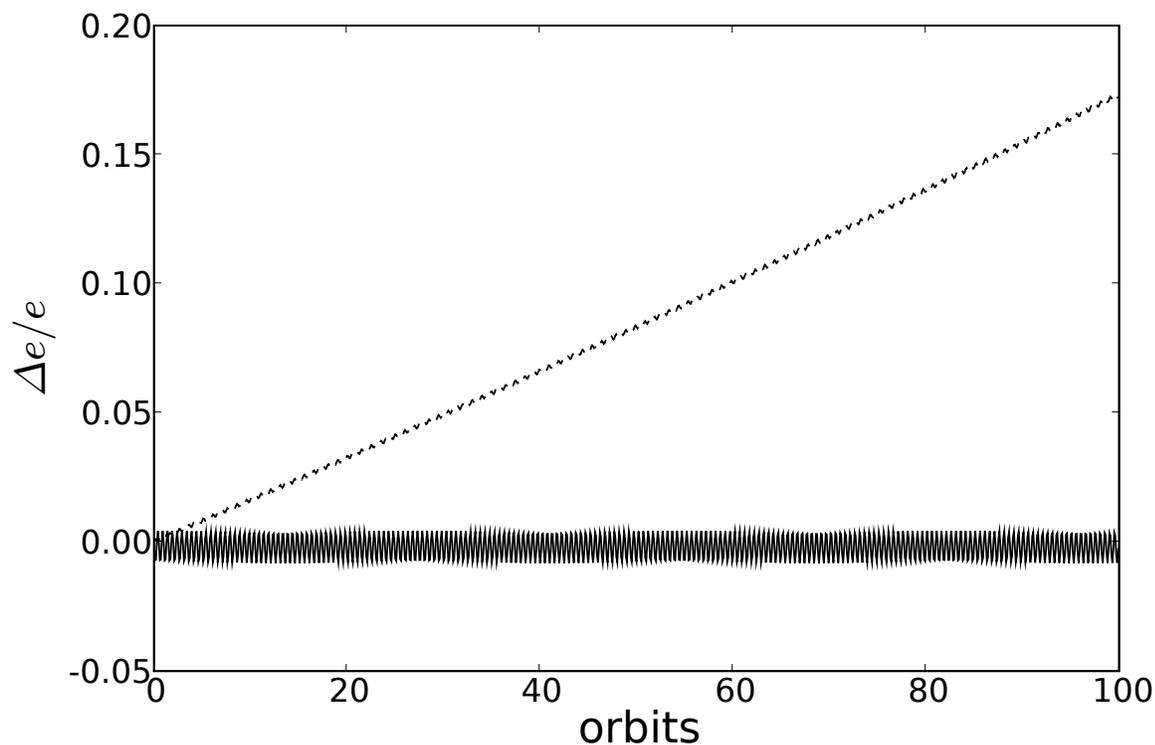}
\caption[]{
\label{evst}
Relative change in eccentricity for a single particle.  $\Delta e/e$
is plotted vs orbit number for a 100-orbit integration of a particle
with an initial eccentricity of 0.001.  The dotted line
shows the evolution using the standard non-symplectic integration
algorithm with 100 steps per orbit.  The solid line shows the evolution
using the symplectic algorithm with 20 steps per orbit.  The
symplectic algorithm with 100 steps per orbit has a maximum $\Delta
e/e$ of 0.00017, so its evolution would be a horizontal line at the
resolution of this plot.  All integrations were performed with {\em PKDGRAV}
on a single processor.
}
\end{figure}

\section{Tests of the Method}

As a first test of the usefulness of a symplectic scheme for Hill's
equations, we used the implementation in {\em PKDGRAV} to integrate a single
particle in a Kepler potential and
compare the conserved integrals of the system using our new
integrator with those using a standard second-order integration
method.  The standard method integrates equations (\ref{eq:motionx})
and (\ref{eq:motiony}) with the {\em Kick-Drift-Kick} leapfrog
described above, with the modification that velocities are predicted
to the end of the timestep using the old accelerations in order to
calculate the velocity dependent part of the force.  This is the same
algorithm that is used to handle the velocity dependent forces arising
in Smooth Particle Hydrodynamics \citep{Wadsley04}.

In this case of a single particle $\Phi = 0$, and
a physically relevant combination of the integrals is the
eccentricity, which in terms of the canonical coordinates can be
expressed as
\begin{equation}
e^2 = {1 \over R^2 \Omega^2}\left(P_x^2 - 4 \Omega x P_y + \Omega^2
x^2 + 4 P_y^2\right),
\end{equation}
where $R$ is the radius of the orbit of the patch.
Figure \ref{evst} shows the relative change in eccentricity in an
integration of a particle with an initial $e = 0.001$ for 100 orbits.
The dotted line shows this change for the standard integration method
when 100 steps per orbit are used, while the solid line shows results
for our new method with only 20 steps per orbit.
The new integrator demonstrates the typical behavior of a symplectic
method: the numerical value of the integral of motion oscillates
around the true value, and there is no secular drift.  An integration
with the symplectic integrator using 100 steps per orbit has a maximum
$\Delta e/e$ of 0.00017.  Comparing this with the maximum $\Delta e/e$ for the
integration with 20 steps per orbit, 0.0043, indicates that the error is
scaling as $\tau^2$, where $\tau$ is the timestep.  This is as
expected for a second-order integrator.  A careful inspection of the
figure shows that the
dotted line has a slope that is slightly increasing with time,
implying that in the standard method, the eccentricity drift in this
case grows faster for larger eccentricities.

%

A somewhat more relevant test of the integrator is following an
encounter in the restricted three body approximation.  Specifically, we use
the {\em PKDGRAV} implementation to 
integrate the orbit of a test particle as it comes within a Hill
radius of a massive
body on a circular orbit.  Such a situation is not uncommon in
simulations of planetesimal growth: the large bodies are in somewhat
circular orbits (BQLR).  The accuracy of the integration can
be evaluated
using the Jacobi integral \citep{DQT89}
\begin{equation}
\Gamma = 3 \Omega^2 x^2 - \dot x^2 - \dot y^2 + { 2 G m \over (x^2 +
  y^2)^{1/2}},
\end{equation}
where $m$ is the mass of the massive body.  Guided by the end-state
configuration of the $L_1$ simulation in BQLR, we set the mass of the massive
body to be $3.78\times 10^{18}\,$g (300 times the mass of a 1 km
planetesimal).  The test particle is placed in an orbit such that it
comes within one Hill radius of the massive body at aphelion, and the
encounter speed at closest approach is given by the RMS velocity in
the BQLR simulation, 2 m s$^{-1}$.  These parameters imply an eccentricity
of $1.6 \times 10^{-4}$ and a relative difference in semi-major axis
$\varepsilon 
\equiv |a - a_m|/a = 1.69 \times 10^{-4}$, where $a_m$ is the
semi-major axis of the 
massive body.  We follow the motion of the test particle
starting at perihelion, through the conjunction with the massive body
and to the subsequent perihelion.  Due to the encounter, the
eccentricity of the test body changes by $6.4 \times 10^{-7}$.  This
is somewhat
greater than the eccentricity change expected from the mapping
formula of \cite{DQT89}, $1.5\times 10^{-7}$, presumably because this
encounter does not
satisfy their approximation that $e \ll \varepsilon$.

\begin{figure}
\plotone{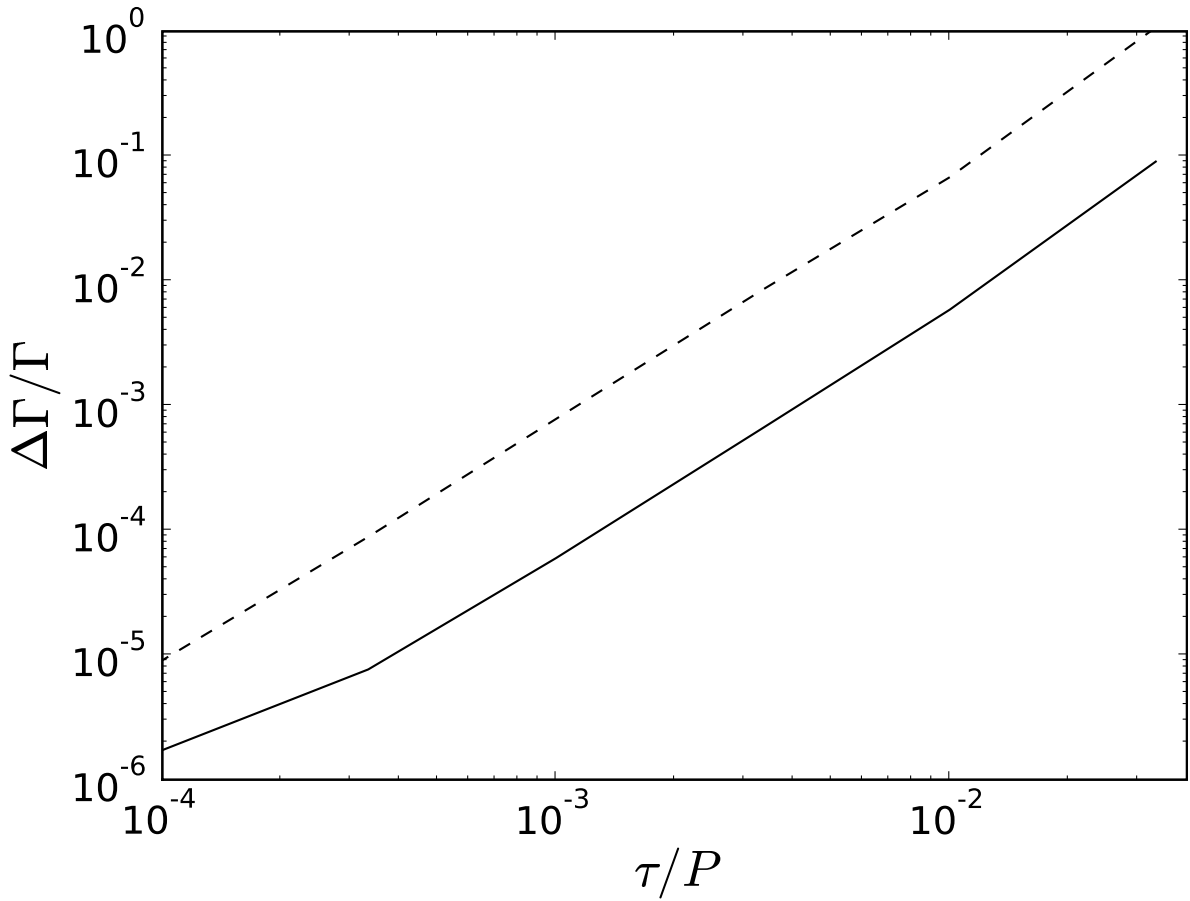}
\caption[]{
\label{encounter}
Maximum relative change in the Jacobi constant, $\Gamma$, during an
encounter with a massive body as a function of step size.  A single
orbit during which the test body comes within one Hill radius of a
massive body is integrated between two successive perihelia.  The
dotted line shows the maximum change in $\Gamma$ over this orbit as a
function of step size in units of the orbital period for
the standard non-symplectic integration.  The solid line shows the
same quantity for the symplectic algorithm.
All integrations were performed with {\em PKDGRAV} on a single processor.
}
\end{figure}

Figure \ref{encounter} shows how well $\Gamma$ is conserved during
this encounter as a function of the integration timestep for the symplectic
and the non-symplectic integrations.  As with the eccentricity in the
simple orbit case, the error in the Jacobi constant scales as
$\tau^2$.  However, for a given timestep the symplectic integration
algorithm gives an order of magnitude improvement in the conservation
of $\Gamma$.  At the largest timestep plotted, $\tau = P/30$, the test
body moves through the encounter at about 2 $r_H$ per timestep.  Hence
reasonably resolving the encounter requires $\tau = P/100$ or smaller.

\section{Discussion and Summary}

The choice of algorithm for a numerical simulation is critical
to obtaining accurate physical results.  This is particularly true for
simulations performed over many dynamical times where there is an
opportunity for truncation error to build up in a secular manner.  In
this case, an algorithm that appears to work well over a few orbits
may produce unacceptably incorrect results when used for hundreds of
orbits.  Using the standard second-order integrator to solve Hill's
equations in the context of planetesimal accretion illustrates this
problem.  Although individual orbits are followed reasonably well with
a few hundred steps per orbit, the secular growth in eccentricity
shown in Fig.~\ref{evst} could overwhelm any physical changes in
the eccentricity distribution of planetesimals in a hundred orbits or
so.  Faced with this problem, one must either go to a higher-order
algorithm, which is difficult to implement in a large parallel
simulation code, or use much smaller timesteps, which significantly
increases the computational expense.

Fortunately, for the case of Hill's equation we have discovered a
second-order symplectic integrator that does not display any secular growth
in eccentricity.  Moreover, our solution is linear in the time
extrapolation of particle positions, permitting efficient collision
detection, which is valuable for most large $N$ simulations of
planetesimals and planetary rings. The algorithm is derived from the
formalism of
\cite{SahaTremaine92} where the Hamiltonian is separated into
parts that by themselves are integrable.  This separation
is in turn made possible via a canonical transformation to coordinates
that significantly simplifies the Hamiltonian.  The algorithm has been
implemented in the scalable parallel $N$-body code {\em PKDGRAV}, and
this code is currently being used for follow-on simulations to those
described in BQLR, which simulated the growth of planetesimals over
hundreds of orbits.

As described above, our integrator has a fixed timestep which is
inefficient for simulations that have to resolve close encounters
between bodies where the encounter timescales are a small fraction of
the orbital time.  For example, in planet formation simulations the
encounter timescale is hours compared to an orbital time of one year.
Constructing an integrator that can adjust
timesteps in order to handle close encounters and yet retains
symplectic properties requires care
\citep{DuncanLevisonLee98,Chambers99}.  In our implementation, we have
simply adjusted the timestep of particles that are experiencing a
collision.  Although this destroys the symplectic properties of the
integrator, for the planetesimal simulations in BQLR, a typical particle
makes about one hundred orbits before experiencing a collision.  Hence,
even with the non-symplectic timestep adjustment, our integrator
significantly improves the quality of these simulations.

\section{Acknowledgments}

This work was supported by NASA's Terrestrial Planet Finder Foundation
Science/Solar Systems Origins program under Grant 811073.02.07.01.15.
Randall Perrine was supported by a NASA Earth and Space Science Fellowship.


\bibliography{trq}

\end{document}